\documentclass{article}
\usepackage{color}
\usepackage[space]{grffile}
\usepackage{changes}
\usepackage{ulem}
\usepackage{authblk}
\usepackage[utf8]{inputenc}
\usepackage{siunitx}
\usepackage{amsfonts}
\usepackage{amssymb}
\usepackage{amsthm}
\usepackage{algorithmic,algorithm}
\usepackage{color}
\usepackage{mathrsfs}
\usepackage{graphicx, float}
\graphicspath{{./Figures/}}
\usepackage{subfigure} 
\usepackage{proba}
\usepackage{booktabs}
\usepackage{textcomp}
\usepackage{longtable}
\usepackage{tabulary}
\usepackage{booktabs,array,multirow}
\usepackage{amsfonts,amsmath,amssymb}

\usepackage{natbib}
\usepackage{url}
\usepackage{hyperref}
\hypersetup{colorlinks=false,pdfborder={0 0 0}}
\usepackage{etoolbox}
\usepackage{cleveref}
\usepackage{siunitx}

\newcommand{\Pb}{\mathbb P}

\newtheorem{theorem}{Theorem}
\newtheorem{hypotheses}[theorem]{Hypotheses}

\newtheorem{rem}{Remark}
\newtheorem{exm}{Example}

\begin{document}
    \author[1]{Fernando Baltazar-Larios}
     \author[2]{Francisco Delgado-Vences}
   \author[3]{Saul Diaz-Infante}
   \affil[1]{\small Departamento de Matem\'aticas, Facultad de Ciencias, Universidad Nacional Aut\'onoma de M\'exico}
     \affil[2]{\small  CONACYT - Instituto de Matem\'aticas, Universidad Nacional Aut\'onoma de M\'exico} 
      \affil[3]{\small  Departamento de Matem\'aticas, %
        CONACYT - Universidad de Sonora, %
        Sonora, %
        Mexico}
    
    \title{
         Maximum likelihood estimation for a % 
    stochastic SEIR system for COVID-19}
   
   \date{Nov 24, 2021}
    \maketitle
   
    \begin{abstract}%
    The parameter estimation of epidemic data-driven models is a crucial task. 
    In some cases, we can formulate a better model by describing uncertainty 
    with appropriate noise terms. However, because of the limited extent and 
    partial information, (in general) this kind of model leads to intractable 
    likelihoods. Here, we illustrate how a stochastic extension of 
    the SEIR model improves the uncertainty quantification of an overestimated
    MCMC scheme based on its deterministic model to count 
    reported-confirmed COVID-19 cases in Mexico City. 
    Using a particular mechanism to manage missing data, we developed MLE 
    for some parameters of the stochastic model, which improves the description 
    of variance of the actual data.\\
 {\bf keywords} :
        COVID-19,
        Stochastic SEIR Model,
        Maximum Likelihood Estimation
    
\end{abstract}   
    
    \section{Introduction}
        
Parameter calibration of epidemiological models with data 
is a demanding but crucial task. 
The current COVID-19 pandemic has dramatically revealed 
the complexities and subtleties of developing epidemic 
data-driven models.
Although computational cost would grow when we develop a
model with stochastic differential equations, in some particular
cases this would lead to tractable estimators. By quantifying uncertainty
by Brownian motion and applying methods based on maximum likelihood,
we can improve the calibration of a deterministic base with conventional 
methods. When we calibrate parameters of an epidemic model, 
we argue that we can sometimes improve the quality
and computing effort by adding noise and proper information management.

    Models based on stochastic differential equations (SDEs)
have the advantage of characterising the central tendencies\textemdash 
usually a regarding deterministic model to describe the mean. 
This stochastic structure could also capture sources of variability
\citep[see e.g.][]{Allen2007,Kloeden2017,Gardiner2009}. 
These models represent an alternative to analyse 
the skewed data that is commonly sampled in the observations. 
Stochastic models, and particularly SDEs, attempt to capture
in their formulation the random variability of the data, and therefore  
quantify the uncertainty that is naturally related to parameters or 
phenomena \cite{Allen2007}.
    
    One of the crucial challenges in likelihood-based inference 
for scalar SDEs is to compute of the underlying transition 
density, and hence the likelihood function. Thus, maximum likelihood 
estimation on the true likelihood of the solution of SDEs is a rare case. 
Moreover, because the available sampled solution 
is a sequence of discrete time observations, it is usually incomplete and 
limited, and we must instead appeal to approximations \cite{iacus}. 
Furthermore, this situation worsens for a coupled system of SDEs.

    Regarding stochastic models for epidemiology, and in particular 
for COVID-19, in previous studies there are two types of works related 
to the one presented here. 
On one hand, some manuscripts report similar ideas but for simpler structures 
and with numeric experiments of synthetic nature.
For example, \citeauthor{PaGrGrMa} reports the estimation of a SIS
structure that only needs a scalar version of the model to
approximate the likelihood \cite{PaGrGrMa}. 
The authors use a pseudo-likelihood and least square method to estimate 
the parameters in the model. They also calculate  confidence intervals by 
applying least squares. 
In contrast, the results presented here deal with an epidemic model 
of five non-linear coupled SDEs and with real data for COVID-19 
from Mexico City, we also have a natural discretised version 
of the MLE to estimate the parameters in the model.  
    
    On the other hand, there are complex stochastic models similar to 
the one considered in this manuscript that are only studied on a 
theoretical basis and/or with numerical simulations. 
However, this does not address estimations methods. 
For example, in \cite{faranda2020modeling}, the authors propose
a SEIR model, with four classes of individuals, given by a 
random differential equation. Indeed, they consider a model where
two parameters are Brownian motion with drift and one parameter is the solution 
of a SDE with a log-normal perturbation. 
Nevertheless, the authors do not estimate parameters 
from actual data to fit the model, the parameters are instead taken from 
previous works. Additionally, they used date information about 
confinement in Italy and France to validate the model.
%---------------------------------------------------------------------------
In \cite{djordjevic2021two}, the authors study a version of the stochastic
SEIR model  for the COVID-19. Their model is a system of stochastic 
differential equations driven by two independent standard Brownian motions. 
More precisely, their model depends on several parameters,
including the constant transmission coefficients related to symptomatic 
infected individuals ($\beta$), hospitalised individuals 
and to superspreaders ($\beta'$). In their study, the authors formulate a
stochastic model via a stochastic perturbation of the constant 
transmission rates $\beta,\beta'$. They show the existence and uniqueness 
of the global positive solution. They also study conditions
under which extinction and persistence in mean hold.  
Finally, they provide  numerical simulations of the stochastic model. 
Nevertheless, estimation methods to fit the model is an open problem for their model.
 
%------------------------------------------------------------------------------
    This manuscript focus on the calibration of a stochastic SEIR
(Susceptible-Exposed-Infective-Recovered) model %\cite{Anderson} 
with data of symptomatic reported and confirmed 
cases in Mexico City. According to a specific time window
\textemdash where the outbreak follows an increasing profile\textemdash, 
our approach relies on the maximum likelihood estimation method.
   
Here, we formulate our stochastic SEIR model
through a stochastic perturbation of the natural death rate of the 
population by a Brownian motion.  We then deduce a nonlinear and 
coupled It\^o stochastic differential equation (SDE), driven by 
an affine Gaussian multiplicative noise.
   
Our main objective pursuit the development of consistent
Maximum Likelihood Estimator (MLE)  for a stochastic 
SEIR structure.

     Parameter estimation for SDEs is an active  
area in the statistical literature. For example, in \citet{Young81}
\citeauthor{Young81} gives an overview of parameter estimation
methods with continuous-time observation.  
\citeauthor*{Soren-04} treat the case where the data correspond
to the solution of SDEs observed in a sequence of discrete time points.  
We refer the reader to \cite{Soren-04} for an introduction of several
inference methods as estimating functions, analytical and numerical 
approximations of the likelihood function, Bayesian analysis Multi Chain
Monte Carlo (MCMC) methods, indirect inference, 
and Expectation-Maximisation (EM) algorithm. 

    Due to the COVID-19 pandemic, the estimation of parameters  
and development of epidemic models are under special attention. The most 
recent advances combine several models and data 
estimation. Among others, we see relevant methods based on Kalman filters,
data aggregation \cite{Fintzi2017}, and techniques from data science such as
Artificial Intelligence and machine learning \cite{Bragazzi2018}.

    Our work is related to \cite{Liu,PaGrGrMa,Ndanguza,Hotta,Rios}.
\citeauthor*{Liu} face the estimation 
for a system of SDEs in \cite{Liu}. \citet{Ndanguza} focus on the 
calibration parameters of alternative stochastic SEIR  versions
by adaptive MCMC and extended Kalman filter methods. \citeauthor*{Hotta}
report other Bayesian techniques in \cite{Hotta}, and \citet{Rios}
apply a maximum likelihood method,  in \cite{Otunuga2021}
\citeauthor*{Otunuga2021} reports a parameter estimation with local 
lagged adapted generalised methods of moments.

    Our contribution consists of the development estimators for some crucial 
parameters of the stochastic SEIR model. Here, we apply the quadratic 
variation of all of thee processes to estimate the volatility parameter 
and calculate the MLE for parameters representing the symptomatic infection
rate, asymptomatic infection rate, and proportion of symptomatic individuals.
Then, from a theoretical result, we deduce a target parameters likelihood 
without assuming a prior data distribution. 
In short, this likelihood and a mechanism to manage incomplete data 
form the core mechanism that allows us to estimate the parameters.

Data-driven epidemic models usually treat observations as counters.
Assuming that data follows a given distribution, this could be a Poisson,
negative binomial, and so on. \cite{ONeill2010}. Then, by applying 
some statistical inference tools to estimate some parameters, 
the model is fitted to the incidence data. 
Here, we avoid this assumption. 
A description of the procedure follows.
We use the Radon-Nikodym derivative of two equivalent 
measures\textemdash which, in our case, are the respective solutions 
of the stochastic SEIR system with two different sets of values of 
the parameters\textemdash to obtain the likelihood of the parameters 
of interest. This means that we use the Cameron-Martin-Girsanov 
Theorem to set up the likelihood.
Note that the method to establish the likelihood does 
not assume a particular distribution of the solution, 
therefore,\textemdash we assume that the data have 
an abstract distribution. 

    We pursue estimators that satisfy strong consistency and
numerical schemes of high order to its computing. 
With these ideas, we also illustrate their efficacy
in the time of computing and performance with real-data 
by simulation.

    Our simulations suggest that the methodology developed 
here improves a particular outcome of a fitting with MCMC. 
The scheme of prior distributions returns a bias fitting 
with this method. In contrast, the fitting of the SDE model
with MLE improves the capture of the data variance and the
time of computing.
    According to our Theorem about consistency, by computing 
residuals we confirm  that the stochastic 
model follows the profile of the regarding data and is
coherent with our modelling assumptions.
    
    The results presented here point to a methodology 
that would boost the uncertainty quantification by SDE 
epidemic models with incomplete information and with 
dramatically less computing time. Here, we report the first 
pieces of evidence.

    After this brief introduction, in 
\Cref{sec:deterministic_base_model} we treat a deterministic SEIR
model for COVID-19 and its calibration with MCMC. 
\Cref{sec:inference} forms the main content of this article:
the construction of estimators for important parameters and the analysis 
of its consistency. We illustrate our methodology via simulation in 
\Cref{sec:simulation,sec:application_real_data} by synthetic generated 
data and data of confirmed symptomatic cases from Mexico City.
    \section{Deterministic COVID-19 base dynamics}\label{sec:deterministic_base_model}
                    We consider that susceptible individuals become 
        infected when they are in contact with asymptomatic individuals or
        individuals with symptoms. We propose that a proportion of 
        asymptomatic individuals have a way to get relief and not die. A 
        proportion of individuals infected with symptoms may die of the 
        disease or may be relieved. We will introduce two types of control, 
        vaccination of asymptomatic infected and susceptible individuals and 
        treatment of individuals infected with symptoms. 
\subsection{Deterministic SEIR Structure}
    We use a common SEIR deterministic structure that has been applied for 
    COVID-19. Our formulation splits the entire population $N$ according to 
    the following compartments:
    \begin{description}
    \item[Suceptible $(S)$]
        This compartment denotes the populations' member under risk of contagious.
    \item[Exposed $(E)$]
        Members of this class come from the susceptible class and represent
        the population fraction that hosts but can not transmit the SARS-CoV-2. 
        Then, after a period of $\kappa^{-1}$ days, they develop the faculty to spread 
        this virus and became infected.
    \item[Asymptomatic Infected $I_a$]
        This class denotes the population fraction that spreads 
        the SARS-CoV-2 but does not develop symptoms or just develop mild symptoms. This class also enclose all non-reported COVID-19 cases.
    \item[Symptomatic Infected $I_s$]
        Represents the COVID-19 reported and confirmed cases that develop symptoms.
    \item[Recovered]
        After a time period of length $\alpha_s^{-1}$, a member of the 
        infected class enters this compartment to become recovered and 
        immune. 
    \item[Death]
        Here we count the cases that lamentably dies due to COVID-19.
    \end{description}
    In symbols, our deterministic base reads
    \begin{equation}
        \label{eqn:base_dynamics}
        \begin{aligned}
            f_{\beta}&:= \beta_s I_S + \beta_a I_a
            \\
            S'  &= 
                \mu +\gamma R 
                - \left(\mu  + f_{\beta} \right)  S
            \\
             E' & =  f_{\beta}S 
                 - (\kappa E + \mu E
                 )
            \\
            {I_a}' &=  
                p \kappa E 
                - \big(\alpha_a + \mu \big)I_a   
            \\
             {I_s}' &= 
                (1 - p) \kappa E 
                - (\alpha_s +\mu)  I_s
            \\
            {R}' &=  
                \alpha_a I_a + \alpha_s (1 - \theta)I_s -
                (\mu + \gamma) R
            \\
            D' &= \theta \alpha_s I_s.
        \end{aligned}
    \end{equation} 
    \Cref{table:parametedescription} enclose a description of 
    the underlying parameters.
    \begin{table*}[tbh]
    	\centering
    	\begin{tabular}{rl	
    	}
    		\toprule
    		Parameter & Description
    		\\
    		\midrule
    		$\mu$ &  
    			Natural death rate
    		\\
    		$\beta_s$ & 
    			 Symptomatic infection rate 
    		\\
    		$\beta_a$ & 
    			Asymptomatic infection rate
    		\\
    		$\kappa$ &  
    			Transfer rate from the exposed class to 
                the infected stage, 
            \\
                &
    			thus $\kappa^{-1}$ is the average incubation time.
    		\\
    		$p$ & 
    			Proportion of asymptomatic individuals  
    		\\			
    		$\theta$ & 
    			Proportion of symptomatic individuals who die due to 
    			the disease 
    		\\
    		$\alpha_{s}^{-1}$
    		& Symptomatic recovery period
    		\\
    		$\alpha_{a}^{-1}$
    		& Asymptomatic recovery average time
    		\\ 
    		$\gamma^{-1}$ 
    		&  Disease-immunity period  
    		\\
    		\bottomrule
    	\end{tabular}
    	\caption{
    	    Parameters definition of the model in 
    	    \Cref{eqn:base_dynamics}.
    	}
    	\label{table:parametedescription}
    \end{table*}
    
        To calibrate parameters of this deterministic base 
    \eqref{eqn:base_dynamics}, we deploy a MCMC by counting 
    the cumulative incidence of new infected and reported 
    symptomatic cases. We denote by $Y_t$ this cumulative 
    incidence at time $t$ and suppose that its profile 
    follows a Poisson distribution with data mean 
    $\lambda_t = \E Y_t$. Thus, following ideas 
    from \cite{Acuna2020} we postulate priors for $p, \beta_s, \beta_a $ 
    and count the cumulative reported-confirmed
    cases in the CDMX-Valle de Mexico database \cite{DataMX}.

        Thus, our Bayesian estimation scheme results
    \begin{equation*}
        \begin{aligned}
            Y_t & \sim 
                \mathrm{Poisson}(\lambda_t)
            \\
                \lambda_t =& \int_0 ^ t (1-p) \delta_E E
            \\
                p & \sim 
                \mathrm{Uniform}(0.3, 08)
            \\
                \kappa & \sim 
                \mathrm{Gamma}(10, 50)
        \end{aligned}
    \end{equation*}
    \Cref{table:parametermodel} 
    display the output of the mentioned MCMCM using the dynamic
    Hamiltonian developed by Betancourt 
    \cite{Hoffman2014, Betancourt2017}
    from the rstan implementation with a sample of \num{10000} 
    paths, see [GITHUB] form more details.
    \begin{table*}[tbh]
    	\centering
    	\begin{tabular}{%w
    			%>{\centering}
    			%p{1in}
    			%p{3in}
    			%p{0.57\textwidth}
    		rlc	
    	}
    		\toprule
    		Parameter & Value & Reference
    		\\
    		\midrule
    		$\mu^{-1}$ 
    		& \SI{70} \times \SI{365}{days}
    		& $\dag$
    		\\
    		$\beta_s$ 
    		& \num{0.058215322606755}
    	    & Estimated
    		\\
    		$\beta_a$ 
    		& \num{0.510968165093383}
    		& Estimated
    		\\
    		$\kappa$ 
    		&  \num{0.196078}
    		&  \cite{Tian2020}
            \\
    		$p$ 
    		& \num{0.585505}
    		& Estimated
    		\\			
    		$\theta$ 
    		& \num{0.11}
    		& $\ddag$
    		\\
    		$\alpha_{s}^{-1}$
    		&\num{0.092507}
    		& $\dag$
    		\\
    		$\alpha_{a}^{-1}$
    		&\num{0.167504}
    		& $\dag$
    		\\ 
    		$\gamma^{-1}$ 
    		&\num{1/365}
    		& $\ddag$
    		\\
    		\bottomrule
    	\end{tabular}
    	\caption{
    	    Parameter values of the model in \Cref{eqn:base_dynamics},
    	    \textsuperscript{$\dag$}:\cite{Acuna2020},
    	    {\textsuperscript{$\ddag$}:\cite{Acuna-Zegarra2021}}.
    	}
    	\label{table:parametermodel}
    \end{table*}
\begin{table*}[tbh]
    	\centering
    	\begin{tabular}{%w
    		rll	
    	}
    		\toprule
    		State & Prior 
    		\\
    		\midrule
        		$E(0)$
        		& 
        		    $ 
        		        \displaystyle
        		        \frac{1}{N} 
        		        \cdot 
        		        \mathrm{uniform}(47, 2100)
        		    $
                & 
                $   
                    \displaystyle
                    \frac{
                        \num{198.504524717486}
                    }{N}
                $
    		\\
            		$I_S(0)$
            		&
            		$   
            		    \displaystyle
            		    \num{74} / N
            		$
    		\\
        		$I_a(0)$
        		& 
        		    $ 
        		        \displaystyle
        		        \frac{1}{N} 
        		        \cdot
        		        \mathrm{uniform}(47, 2100)
        		    $
                & 
                    $
                        \displaystyle
                        \frac{
                            \num{99.174034301964}
                        }{N}
                    $
    		\\
        		$R(0)$
        		&
    		    \num{0.0}
    		\\
    		    $S(0)$
    		    &
    		    $1 - (E(0) + I_a(0) + I_s(0) + R(0))$
    		    & 
                    $
                        \displaystyle
                        \frac{
                            \num{26446062.901024}
        		        }{N}
        		    $
    		\\
    		\bottomrule
    	\end{tabular}
    	\caption{Initial conditions of the system of equations (\ref{eqn:base_dynamics}).}
    	\label{table:initial_det}
    \end{table*}
\begin{figure*}[htb]
    \centering
    \includegraphics[width=0.8\textwidth, keepaspectratio]{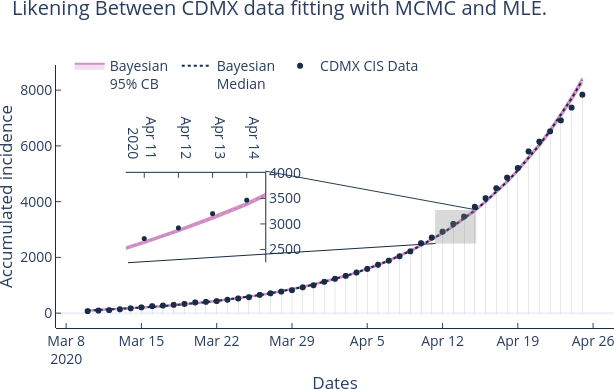}
    \caption{%
        MCMC fit of diary new cases of Mexico city
        during exponential growth. See 
        \url{https://plotly.com/~sauld/53/} for an electronic
        version.
    }
    \label{fig:data_CDMX_fitting}
\end{figure*}
    
    \section{Stochastic extension}
           
        In this section, we consider a stochastic 
    version of the model given by \eqref{eqn:base_dynamics}.
    We aim to quantify uncertainty. To this end, we conveniently 
    perturb the natural mortality rate $\mu$ by a standard Wiener 
    process $W(t)$. That is, we add a stochastic infinitesimal 
    fluctuation of the parameter $\mu$ in the interval
    $(t, t+dt)$ by adding the corresponding Wiener 
    increment $dW(t)$. 
    In symbols, we have 
    \begin{equation}
        \label{eqn:mu_perturbation}
        \mu dt\rightsquigarrow \mu dt + \sigma dW(t).
    \end{equation}
     This perturbation infinitesimally quantifies the environmental 
     fluctuations of the population mortality, due, for example, 
     to differences in the health of individual conditions and 
     other-regarding environmental factors. 
     Furthermore, this particular form allows us to deduce a closed 
     form for the underlying likelihood. Because our model 
     couples the spread and other processes, this perturbation 
     also implicitly captures random fluctuations from all involved
     compartments.
    
        In other words, we describe this effect by a stochastic 
    process with mean zero and dispersion proportional to  
    $\sigma \sqrt{t}$. According to Wiener statistical 
    properties, we have $\EX{\mu dt+\sigma dW(t)}{} = \mu dt$ and
    $\VarX{\mu dt+\sigma dW(t)}{} = \sigma^2 dt$.
    This approach has been employed in the emergent research
    of mathematical epidemiology, see for example \cite{Chang2019, Acunya2018, Liu2018}.
\subsection{Stochastic Perturbation}
     Substituting perturbation \eqref{eqn:mu_perturbation} 
     in equation
     \eqref{eqn:base_dynamics} results,
     \begin{equation}
         \label{eqn:sde_model}
         \begin{aligned}
             d {S}(t)  =& 
                 \big[\mu - \mu S(t) - f_{\beta} S(t) 
                 +\gamma R(t)  \big] dt + \sigma \big(1- S(t)\big) dW(t)
             \\
              d {E}(t) =& \big[  f_{\beta} S(t)
                  - \kappa  E(t) - \mu E(t) \big] dt  - \sigma E(t) dW(t)
             \\
             d {I_a}(t) =& \big[ 
                 p \kappa E(t) 
                 -   (\alpha_a +\mu) I_a(t)  \big] dt  - \sigma I_a(t) dW(t) 
             \\
              d {I_s}(t) =& \big[ 
                 (1 - p) \kappa E (t)
                 - (\alpha_s +\mu)  I_s(t) \big] dt   - \sigma I_s(t) dW(t)  
             \\
             d {R}(t) =&  
                \big[ 
                    \alpha_a I_a(t) + 
                    \alpha_s I_s(t) - (\mu + \gamma) R(t) 
                \big] dt 
                - \sigma R(t) dW(t),
             \qquad
             t \in & [0, T] .
         \end{aligned}
     \end{equation}
Since $dN(t) = dS(t) + dE(t) + dI_a(t) +dI_s(t) + dR(t) = 0$,
normalising with respect to the size of the population $N$, we see that
$$
     S(t) + E(t) + I_a(t) + I_s(t) + R(t) = 1.
$$
    Existence and uniqueness of the solution in a compact time interval $[0,T]$ 
follows from  the boundedness, local Lipschitz and linear growth conditions on
the coefficients of each SDE component (see\cite{Oksendal2003} Theorem 5.2.1).
Furthermore, Burkholder-Davis-Gundy  and Holder inequalities imply that, for 
   $r \geq 2$
 
  $$
       \EX{
            |S(t)| ^ r + |E(t)| ^ r+  |I_a(t)| ^ r + |I_s(t)| ^ r + |R(t)| ^ r
        }{} 
              <C (1 + T + T ^ {r/2} )
       < +\infty,
   $$
  with constant $C>0$ depending on the model parameters. 
  We rewrite the SDEs in \Cref{eqn:sde_model} by applying Lamperti's 
  transformation to each component (see \cite{iacus}). Thus,
  the susceptible component $S$ becomes
  \begin{equation}
    \label{dlogS}
    \begin{aligned}
        -\frac{1}{\sigma}d
        \big(
            \log (1-S_t(t)) 
        \big)   \Big[
                \frac{\mu}{\sigma }
                - \frac{ S(t) f_{\beta}}{\sigma \big(1-S(t)\big)}
                + \frac{\gamma  R(t)}{\sigma \big(1-S(t)\big)} 
                + \tfrac{1}{2} \sigma \Big] dt + dW(t). 
     \end{aligned}
 \end{equation}
    Again, by  Lamperti's transformation for the equation of process $E$, 
    we obtain
\begin{align}
    -\frac{1}{\sigma} 
        d(\log \big(E(t)\big))&= 
            \Big[
            -\frac{  S(t) f_{\beta} }{\sigma E(t)} + 
            \frac{ \kappa }{\sigma } + 
            \frac{\mu }{\sigma } +
            \frac{1}{2} \sigma\Big] dt +
            dW(t).\label{dlogE}
\end{align}
 Analogously, we obtain the Lamperti transformation, for the components processes 
 $I_a,I_s,R$:
 \begin{align}
    -\frac{1}{\sigma} 
        d\big(\log (I_a(t))\big)
        &= 
        \Big[ 
            -\frac{\kappa p E(t)}{\sigma I_a(t)} + 
            \frac{(\alpha_a + \mu )}{\sigma } +
            \frac{1}{2} \sigma\Big] dt + dW(t), \label{dlogIa}
\end{align}
\begin{align}
    -\frac{1}{\sigma}
        d\big(\log (I_s(t))\big)
            &= 
            \Big[
                -\frac{\kappa (1-p) E(t)}{\sigma I_s(t)} +
                \frac{(\alpha_s + \mu )}{\sigma } +  
                \frac{1}{2} \sigma\Big] dt + dW(t), \label{dlogIs}
\end{align}
\begin{align}
    \label{dlogR}
    -\frac{1}{\sigma}
        d\big(\log (R(t))\big)
        &= 
            \Big[ 
                - \frac{\alpha_a I_a(t) + 
                \alpha_s I_s(t)}{\sigma R(t)} + 
                \frac{\mu + \gamma}{\sigma } +  
                \frac{1}{2} \sigma
            \Big] dt 
            + dW(t).
\end{align}

Then, we can write these SDEs as a five dimensions system of SDEs:
\begin{equation}\label{dlog-SE}
    -\frac{1}{\sigma} 
        d \mathbf{X}_{\beta,p}(t) = 
            F\big(\mathbf{X}_{\beta,p}(t)\big) dt + d\mathbf{W}(t),
\end{equation}  
where, 
\begin{align*}
 \mathbf{X}_{\beta,p}(t) &:= 
    \begin{pmatrix}
        \log \big(1-S(t)\big)
        \\ 
            \log \big(E(t)\big) 
        \\
            \log \big(I_a(t)\big)
        \\  
            \log \big(I_s(t)\big) 
        \\  
            \log\big (R(t)\big)
    \end{pmatrix},
    \quad 
    F\big(\mathbf{X}_{\beta,p}(t)\big):=  
    \begin{pmatrix}
        \dfrac{\mu}{\sigma } -
        \dfrac{f_\beta S(t) }{\sigma \big(1-S(t)\big)} 
        + 
        \dfrac{\gamma R(t)}{\sigma \big(1-S(t)\big)} 
        + \tfrac{1}{2} \sigma  
        \\ 
            - \dfrac{f_\beta  S(t)  }{\sigma E(t)} 
            + \dfrac{ \kappa   }{\sigma }  
            + \dfrac{\mu }{\sigma } 
            + \dfrac{1}{2} \sigma 
        \\
            - \dfrac{\kappa p E(t)}{\sigma I_a(t)} 
            + \dfrac{(\alpha_a + \mu )}{\sigma }  
            + \dfrac{1}{2} \sigma 
        \\
            - \dfrac{\kappa (1-p) E(t)}{\sigma I_s(t)} 
            + \dfrac{(\alpha_s + \mu )}{\sigma }  
            + \dfrac{1}{2} \sigma 
        \\
            - \dfrac{\alpha_a I_a(t) + \alpha_s I_s(t)}{\sigma R(t)} 
            + \dfrac{\mu + \gamma}{\sigma } 
            + \dfrac{1}{2} \sigma
    \end{pmatrix},
    \\  
    d\mathbf{W}(t)&:=
    \begin{pmatrix}
        dW(t) 
        \\ 
        dW(t) 
        \\ 
        dW(t)
        \\ 
        dW(t) 
        \\ 
        dW(t)
    \end{pmatrix} .
\end{align*}    
    %\section{Noise calibration with MLE} 
    \section{Statistical inference for $\beta$, $p$ and $\sigma$}\label{sec:inference}
          We apply the version of the  Girsanov formula presented in 
\cite[see Theorem 7.4]{saso19} for the system in
\Cref{dlog-SE}. We assume that the true value of the parameters
$\beta_{s,0},\beta_{a,0}, p_0$ are unknown. We denote by $\Pb_{\beta,p}$ the law for
the solution of \eqref{dlog-SE}. It is known that the measures 
$\Pb_{\beta,p}$ are equivalent for different values of ${\beta,p}$
(see for instance \cite{iacus}, \cite{Bishwal2008}, \cite{saso19}. 
Then, by Theorem 7.4 in \cite {saso19}, we have that the likelihood ratio, 
or Radon-Nikodyn derivative has the form 
\begin{equation} 
    \label{Likelihood}
    \begin{aligned}
        \frac{d\Pb_\beta}{d\Pb_{\beta_0}} 
            &= 
            \exp\Bigg[ \int_0^T [ F(\mathbf{X}_{\beta,p}(t))  - 
            F(\mathbf{X}_{\beta_0,p_0}(t)) ]^{T} Q^{-1} d\mathbf{W}(t) 
            \nonumber\\
                &\qquad \quad -\frac{1}{2} 
                \int_0^T  [ F(\mathbf{X}_{\beta,p}(t))  - 
                F(\mathbf{X}_{\beta_0,p_0}(t)) ]^{T}Q^{-1}  
                [ F(\mathbf{X}_{\beta,p}(t))  - 
                F(\mathbf{X}_{\beta_0,p_0}(t)) ] d(t)\Bigg], 
    \end{aligned}
\end{equation}
where $Q$, in our case, is the five-identity matrix and 
\begin{align*}
    [ F(\mathbf{X}_{\beta,p}(t))  - F(\mathbf{X}_{\beta_0,p_0}(t)) ]  
    &= 
    \begin{pmatrix}
        -(f_\beta-f_{\beta_0}) \dfrac{S(t) }{\sigma (1-S(t))} 
        \\ 
        -(f_\beta-f_{\beta_0}) \dfrac{S(t) }{\sigma E(t)} 
        \\
        -(p - p_0) \dfrac{\kappa E(t)}{\sigma I_a(t)}   
        \\
        (p-p_0) \dfrac{\kappa E(t)}{\sigma I_s(t)}   
        \\   
        0
    \end{pmatrix},
\end{align*}
 with 
 $
    f_\beta-f_{\beta_0} = (\beta_s I_s(t) + \beta_a I_a(t) ) 
        -  (\beta_{s,0} I_s(t) + \beta_{a,0} I_a(t) ) 
        = 
            (\beta_s  -  \beta_{s,0}) I_s(t) 
            + (\beta_a - \beta_{a,0}) I_a(t)  
$.
 
   Note that the MLE for $\beta_a$ and $\beta_s$ is
coupled, but independent of $p$. Therefore, we calculate the
MLE for $\beta_a$ and $\beta_s$ together and for $p$ separately.
 
    Thus,

    \begin{align*}
    [F(\mathbf{X}_{\beta,p}(t)) - F(\mathbf{X}_{\beta_0,p_0}(t)) 
    ]^{T} Q^{-1} d\mathbf{W}(t)  
    &=  -(f_\beta-f_{\beta_0})[
        \dfrac{S(t)}{\sigma (1 - S(t)) }
        dW(t) 
        +
        \dfrac{S(t)}{\sigma E(t)} dW(t) ]\\
    &\quad    - (p-p_0)[
        \dfrac{\kappa E(t)}{\sigma I_a(t)} dW(t) 
        + 
        \dfrac{\kappa E(t)}{\sigma I_s(t)} dW(t)] ,
\end{align*}

and

\begin{align*}
[F(\mathbf{X}_{\beta,p}(t)) - F(\mathbf{X}_{\beta_0,p_0}(t)) ]^{T}Q^{-1}   [F(\mathbf{X}_{\beta,p}(t))  - F(\mathbf{X}_{\beta_0,p_0}(t)) ]  
&= (f_\beta-f_{\beta_0})^2\Big[ \Big(\frac{S(t)}{\sigma (1-S(t))}\Big)^2+ \Big(\frac{S(t)}{\sigma E(t)}\Big)^2\Big]\\
&\quad+(p-p_0)^2 \Big[\Big(\frac{\kappa E(t)}{\sigma I_a(t)} \Big)^2 
+\Big(\frac{\kappa E(t)}{\sigma I_s(t)}\Big)^2\Big].
\end{align*}

    We denote by $\star=a$ or $s$, thus $\beta_\star$ will mean
the variable $\beta_s$  or $\beta_a$ and similarly $I_\star$
means $I_a$ or $I_s$. In addition, when, for instance, $\star=a$
then $\star^c=s$ and vice versa.
%\fer{Deriving with respect to x and equating to zero we have that y omitir la primera ecuacion}

Then,  by maximising the log
likelihood with respect to $\beta_\star$ we have that 

 \begin{align*}
    0 
    &= 
        \int_0^T 
            \Big[ 
                - \frac{S(t) I_\star(t)}{\sigma (1-S(t))} 
                - \frac{S(t) I_\star(t)}{\sigma E(t)} \Big]  
                dW(t)  
        \\
    & - 
        \int_0^T  
        \Big[
            (\beta\star-\beta_{\star,0} )I_\star(t) 
            + (\beta{\star^c} - \beta_{\star^c,0} ) I_{\star^c}(t)
        \Big]I_\star(t) 
        \left(
                \Big[
                    \frac{S(t) }{\sigma (1-S(t))} 
                \Big] ^ 2  
            +
            \Big[ 
                \frac{S(t)}{\sigma E(t)} 
            \Big]^2
        \right) dt.
\end{align*}
then, 
\begin{align*}
    &(\beta\star-\beta_{\star,0} ) 
    \int_0^T 
        \left(  
            \Big[ 
                \frac{S(t) I_\star(t)}{ (1-S(t))} 
            \Big]^2  
            +   
            \Big[ 
                \frac{S(t) I_\star(t)}{ E(t)} 
            \Big]^2 
        \right) dt
    \\
        &\quad 
        + 
        (\beta_{\star^c} - \beta_{\star^c,0} )  
        \int_0 ^ T 
            I_{\star^c}(t) I_\star(t) 
            \left(
                \Big[
                    \frac{S(t) }{ (1-S(t))}
                \Big]^2  
                +
                \Big[ 
                    \frac{S(t) }{ E(t)} 
                \Big]^2 
            \right) dt
    \\ 
        & =    
            - \sigma
            \int_0^T 
                \Big[
                     \frac{S(t) I_\star(t)}{ (1-S(t))} 
                     +  \frac{S(t) I_\star(t)}{ E(t)}
                \Big]  dW(t).
\end{align*}
   From this expression we obtain that 
   %\fer{omitir la primera ecuación}
\begin{equation}
    \label{MLE-beta-beta0_1}
    \begin{pmatrix}
        J_s(T) & J_{sa}(T)\\
        J_{sa}(T & J_a(T)
    \end{pmatrix}
    \begin{pmatrix}
        \hat{\beta_s}  - \beta_{s,0} 
        \\
        \hat{\beta_a} - \beta_{a,0} 
    \end{pmatrix}
    =
        - \sigma \begin{pmatrix} 
        \int_0^T 
            \Big[ 
                \dfrac{S(t) I_s(t)}{ (1-S(t))} 
                + \dfrac{S(t) I_s(t)}{ E(t)}
            \Big] dW(t) 
        \\
        \int_0^T 
            \Big[ 
                \dfrac{S(t) I_a(t)}{(1 - S(t))}
                + \dfrac{S(t) I_a(t)}{E(t)} 
            \Big] 
            dW(t)
    \end{pmatrix},
\end{equation}
where 
\begin{align*}
    J_\star(T)
        &:= 
            \int_0^T 
                \left(
                    \Big[
                        \dfrac{S(t) I_\star(t)}{ (1-S(t))} 
                    \Big]^2  
                    +
                    \Big[ 
                        \dfrac{S(t) I_\star(t)}{ E(t)}
                    \Big]^2 
                \right) dt,
        \\
    J_{sa}(T)
        &:=  
            \int_0^T 
                I_a(t) I_s(t) 
                \left(
                    \Big[
                        \dfrac{S(t)}{(1-S(t))}
                    \Big] ^ 2  
                    +
                    \Big[
                        \dfrac{S(t)}{E(t)} 
                    \Big] ^ 2 
                \right) dt.
\end{align*}
    
    Therefore,
\begin{equation}
    \label{MLE-beta-beta0}
    \begin{pmatrix}
        \hat{\beta_s}  - \beta_{s,0} 
        \\
        \hat{\beta_a} - \beta_{a,0} 
    \end{pmatrix}
    =
    - \sigma 
    \begin{pmatrix}
        J_s(T) & J_{sa}(T)\\
        J_{sa}(T & J_a(T)
    \end{pmatrix}^{-1} \, 
    \begin{pmatrix}
        \int_0^T 
            \Big[ 
                \dfrac{S(t) I_s(t)}{ (1-S(t))} 
                + \dfrac{S(t) I_s(t)}{ E(t)} 
            \Big] 
            dW(t) 
        \\
        \int_0^T 
            \Big[
                \dfrac{S(t) I_a(t)}{ (1-S(t))}
                + \dfrac{S(t) I_a(t)}{ E(t)} 
            \Big]  dW(t)
    \end{pmatrix}.
\end{equation}
   
   Meanwhile, deriving the log likelihood with 
respect to $p$ yields  
\begin{align*}
    (p-p_0)
    \left( 
        \int_0^T  
            \Big[ 
                \dfrac{\kappa^2 E^2(t)}{ I_s^2(t)} 
                + \dfrac{\kappa^2 E^2(t)}{ I_a^2(t)} 
            \Big] dt 
    \right) 
    &= 
        \sigma 
        \int_0^T 
            \Big[
                -\dfrac{\kappa E(t)}{ I_a(t)} +
                \dfrac{\kappa E(t)}{ I_s(t)} \Big] dW(t).
  \end{align*}     
This implies that     
\begin{align}
    \label{MLE_p}
    \hat{p}_{ML}-p_0 &=
        \frac{\sigma }{J_2(T)} 
            \int_0^T 
                \Big[
                    - \dfrac{\kappa E(t)}{ I_a(t)} 
                    + \dfrac{\kappa E(t)}{ I_s(t)} 
                \Big] dW(t),
\end{align}
with 
$$
    J_2(T):=
        \int_0^T
            \Big[
                \dfrac{\kappa ^ 2 E ^ 2(t)}{I_s^2(t)} 
                + \dfrac{\kappa ^ 2 E ^ 2(t)}{ I_a^2(t)}
            \Big] dt.
$$
  
Let  
 $
     X_0 ^ + :=  
        \{
            (
                S(t_0),
                E(t_0),
                I_a(t_0),
                I_s(t_0)
            ) 
        \} 
 $ initial state where all populations classes are strictly positive. Denote by
 $
    \varphi:= 
        \left\{
            \mu, 
            \beta_s, 
            \beta_a, 
            \kappa,
            p ,
            \theta,
            \alpha_s ,
            \alpha_a,
            \gamma 
        \right\},
 $
 a model parameter configuration. Accordingly, to
 the van den Driessche definition, the reproductive number
 for the deterministic version of our dynamics
 results
 $$
    \mathcal{R}_0^{D}:= 
        \dfrac{p \kappa \beta_s }{(\mu + \kappa) (\mu + \alpha_s)}
        +
        \dfrac{(1 - p) \kappa \beta_a}{(\mu + \kappa)(\mu + \alpha_a)}.
 $$
 Thus given the initial condition $X_0 ^ + $ and a parameter configuration
 $\varphi$, we define the set 
 $\Omega ^* = \Omega (X_0 ^ + , \varphi, T^*)$ by
 \begin{equation*}
     \begin{aligned}
        \Omega^{*} :=
        \left \{
            (S, E, I_a, I_s, R) \times [t_0 , T]:
        \right.
                &
            S(T) \leq S(t) < S(t_0),
            \\
            &
        \left.
                E(t) > E(t_0), \ 
                I_a(t) > I_a(t_0), \ 
                I_s(t) > I_s(t_0)
        \right \}.
     \end{aligned}
 \end{equation*}
 To prove consistency and asymptotic normality of the parameter, we will make the following hypotheses.
\begin{hypotheses} \label{set_hypotheses}
    There exists $T_0>0$ such that for all $t\in [0,T_0]$
\end{hypotheses}
    \begin{enumerate}%[ i.]
       \item
           $\mathcal{R}_0 ^ D > 1$
        \item
           The initial condition $X_0 ^ +$ and parameter configuration 
           $\varphi$ are such that ${\Omega^* \neq  \emptyset}$
    \end{enumerate}  
This implies that our results are only valid in a sufficiently large time window.
\begin{rem}
     In other words, Hypothesis 
     \ref{set_hypotheses} assures conditions to estimate parameters 
     in a growth phase of the outbreak. Under these conditions, the 
     susceptible population decreases while the Exposed and Infected classes 
     are in the growth phase. Furthermore, according to the definition of set $\Omega ^ {*}$, this growth phase occurs in a time window of size 
     $(T^* - t_0)$. 
\end{rem}
 
    Hypotheses \ref{set_hypotheses} implies that
\begin{align*}
    \dfrac{1}{1-S(t)} < \dfrac{1}{1-S(0)},
    \qquad
    \dfrac{1}{E(t)} < \frac{1}{E(0)}
    \qquad \forall t\in [0, T_0].
\end{align*}  
Now we prove the consistency of our estimators.
\begin{theorem}\label{Consitency-th}
    Assume that Hypotheses \ref{set_hypotheses} are satisfied.
    The estimators 
    $(\hat{\beta}_{s,ML},\hat{\beta}_{a,ML},\hat{p}_{ML})$ 
    are  strongly consistent; that is,
    \begin{equation}
        \label{consistency}
        \lim_{T \rightarrow T_0} 
        \begin{pmatrix}
            \hat \beta_{s,ML} 
            \\
            \hat \beta_{a,ML} 
            \\
            \hat p_{ML}
        \end{pmatrix}
        = 
        \begin{pmatrix}
            \beta_{s,0} 
        \\
            \beta_{a,0} 
        \\
            p_0
        \end{pmatrix}, 
        \qquad \mbox{with probability one.}
    \end{equation}
\end{theorem}
The proof of this theorem is deferred to supplementary
material.

To estimate the parameter of diffusion $\sigma$, 
we can use the quadratic variation over $[0,T]$ of the processes of system of SDEs denoted by $<*,*>_T$. The estimator 
is given by
\begin{equation}\label{est_sigma}
    \hat{\sigma}^2:=\sum_{i=1}^5\frac{\hat{\sigma}^2_i}{5},
\end{equation}
where
\begin{equation}
    \label{sigmas}
    \begin{aligned}
        &\hat{\sigma}^2_1
             = \frac{<S,S>_T}{\int_0^T(1-S(t))^2dt},
        \qquad
          \hat{\sigma}^2_2
             = \frac{<E,E>_T}{\int_0^TE(t)^2dt},
        % %
         \qquad
          \hat{\sigma}^2_3
             = \frac{<I_a,I_a>_T}{\int_0^TI_a(t)^2dt},
        % %
         \\
         & \hat{\sigma}^2_4
             = \frac{<I_s,I_s>_T}{\int_0^TI_s(t)^2dt},
        % %
         \qquad
          \hat{\sigma}^2_5
             = \frac{<R,R>_T}{\int_0^TR(t)^2dt}.
    \end{aligned}  
\end{equation}

     \section{Simulation study}\label{sec:simulation}
     Here, we use the Milstein scheme \cite{Kloeden1992} 
to generate paths of processes of system \eqref{eqn:sde_model} 
in a time interval $[0,T]$. Then we apply \eqref{est_sigma} 
to estimate $\sigma$ and expressions in
\cref{beta_hat_f,beta_hat_a_f,p_hat} to compute the MLE of $\beta_s,\beta_a$, 
and $p$. 
\\
Using the Milstein scheme, we divide the interval $[0,T]$ into $n$ sub-intervals of length 
$\Delta=\frac{T}{n}$ and we obtain observations of each process at the times 
$0=t_0<t_1<\ldots<t_n=T$ where $t_i-t_{i-1}=\Delta, i=1,\ldots,n$. Based on 
these observations the diffusion parameter $\sigma$ can be estimated using the expression 
(\ref{est_sigma}) where

\begin{equation*}
    \begin{aligned}
    \hat{\sigma}^2_1
        &=
        \dfrac{
            2 \sum_{i=1}^{n}
                (S(t_i)-S(t_{i-1}))^2
        }{
            \Delta
            \sum_{i=1}^{n}
                (1-S(t_i))^2+(1-S(t_{i-1}))^2
        },
    \qquad
    \hat{\sigma}^2_2 =
        \dfrac{
            2 \sum_{i=1}^{n}
                (E(t_i)-E(t_{i-1}))^2
        }{
            \Delta
            \sum_{i=1}^{n}
                (
                    E(t_i))^2+(E(t_{i-1})
                ) ^ 2},
    \\
    \hat{\sigma}^2_3
        &=
        \dfrac{
            2 \sum_{i=1}^{n}
                (I_a(t_i)-I_a(t_{i-1}))^2
        }{
            \Delta \sum_{i=1}^{n}
                (I_a(t_i))^2 + (I_a(t_{i-1}))^2
        },
    \qquad
    \hat{\sigma}^2_4
        =\dfrac{
            2\sum_{i=1}^{n}
                (I_s(t_i)-I_s(t_{i-1}))^2
            }{
                \Delta\sum_{i=1}^{n}
                    (I_s(t_i))^2 + (I_s(t_{i-1})
                )^2}, \\
     \hat{\sigma}^2_5
        & =
        \dfrac{
            2\sum_{i=1}^{n}
                (R(t_i)-R(t_{i-1}))^2
        }{
            \Delta \sum_{i=1}^{n}
            (R(t_i))^2+(R(t_{i-1}))^2
        }.
   \end{aligned}  
\end{equation*}
Since
    $$
    \frac{(\hat{\sigma}^2)-\sigma^2}{\sqrt{\Delta}}
        \to
        \mathcal{N}
        \Big(0,\frac{2\sigma^4}{T}\Big),
        \hspace{1cm}\mbox{as}\hspace{.5cm} 
        \Delta\rightarrow 0,
    $$
we deduce that $\hat{\sigma}^2$   
is an unbiased estimator of $\sigma^2$ \citep[see][]{Wei2006}.

Meanwhile, to obtain the MLE of $\beta_a$, $\beta_s$, and $p$, we use the discrete time version 
of expressions (\ref{beta_hat_f}), (\ref{beta_hat_a_f}) and (\ref{p_hat}).
Now, we present two examples with different parameters to be estimated  and initial conditions. In both examples, we use the parameters of Table \ref{table:parametermodel} for $\mu,\kappa,\alpha_a,\alpha_s$ and $\gamma$.
\begin{exm}
    We simulate a \num{1000} datasets in the time interval $[0,1]$. 
    To generate the datasets we use 
    $   \beta_a=0.251521
    $,
    $
        \beta_s=0.45639
    $,
    $
        p=0.1213
    $,
    $
        \sigma=1/5000
    $
     and $\Delta=1/850$ for the approximation in 
    Milstein method and the initial conditions from  
    \Cref{table:initial1}.  
\end{exm}
\begin{table*}[tbh]
    	\centering
    	\begin{tabular}{
    		rl	
    	}
    		\toprule
    		Process & Value
    		\\
    		\midrule
    		$S(0)$ &  $1-(E(0)+I_a(0)+I_s(0)+R(0))$ 
    			
    		\\
    		$E(0)$ & \num{1E-007}
    			 
    		\\
    		$I_a(0)$ & 
    			\num{1E-010}
    		\\
    		$I_s(0)$ & \num{5E-007}
    			
            \\
               
    		$R(0)$ & 
    			\num{0} 
    		\\			
    	
    		\bottomrule
    	\end{tabular}
    	\caption{Initial Conditions for \eqref{eqn:mu_perturbation}.}
    	\label{table:initial1}
    \end{table*}
Using the observation at times $t_0=0,t_1=\Delta,t_2=2\Delta,\ldots,t_{850}=1$, the values of Table \ref{table:parametermodel} for $\mu,\kappa,\alpha_a,\alpha_s,\gamma$ and $\Delta=1/850$, and the discrete version of expressions (\ref{est_sigma}), (\ref{beta_hat_f}), (\ref{beta_hat_a_f}) and (\ref{p_hat}) we find the corresponding estimators.\\
\\
The average of the $1,000$ datasets and the corresponding standard deviations are presented in Table \ref{table:MLE}.

    \begin{table*}[tbh]
    	\centering
    	\begin{tabular}{%w
    			%>{\centering}
    			%p{1in}
    			%p{3in}
    			%p{0.57\textwidth}
    		rlcc
    	}
    		\toprule
    		Parameter & Real Value & Estimator & SD
    		\\
    		\midrule
    		
    		$\beta_s$ 
    		& \num{0.45639}
    	    & \num{0.45315558242869264} & \num{2.9005497641652361E-005}
    		\\
    		$\beta_a$ 
    		& \num{0.25152}
    		& \num{0.26522878914131477} & \num{6.0813886730227286E-002 } 
    		\\
    	
    		$p$ 
    		& \num{0.12130}
    		&\num{0.12035408111700310} & \num{2.5546046918603083E-006}
    		\\			
    		$\sigma$ & \num{2E-004}   & \num{1.3611770138548327E-004}
    		& \num{5.3258165404621520E-009}\\
    		\bottomrule
    	\end{tabular}
    	\caption{Average and  Standard Deviation (SD) of estimator of $\sigma$ and MLEs for $\beta_a,\beta_s$, and $p$.}
    	\label{table:MLE}
    \end{table*}

\begin{exm}
    We simulate  \num{1000} datasets in the time interval $[0,1]$. 
    To generate the datasets, we use the parameters 
    $\Delta=1/1000$, 
    $
        \gamma=1/365,
        \kappa = \num{0.196078},
     \alpha_a = \num{0.167504},
        \alpha_s = \num{0.0925069}
    $ and $N=\num{26446435}$ for the approximation in  Milstein method and the initial conditions from Table \ref{table:initial}.  
\end{exm}

\begin{table*}[tbh]
    	\centering
    	\begin{tabular}{
    		rl	
    	}
    		\toprule
    		Process & Value
    		\\
    		\midrule
    		
    		$E(0)$ & \num{198.504524717486}/N
    			 
    		\\
    		$I_a(0)$ & 
    			\num{99.174034301964}/N
    		\\
    		$I_s(0)$ & \num{74}/N
    			
            \\
               
    		$R(0)$ & 
    			\num{0} 
    		\\
    		$S(0)$ &  $1-(E(0)+I_a(0)+I_s(0)+R(0))$ 
    	    \\		
    	
    		\bottomrule
    	\end{tabular}
    	\caption{Initial Conditions for \eqref{eqn:mu_perturbation}.}
    	\label{table:initial}
    \end{table*}
    
  The average of the $1,000$ datasets and the corresponding 95\% quantiles are presented in Table \ref{table:MLE2}. 
  
  \begin{table*}[tbh]
    	\centering
    	\begin{tabular}{%w
    			%>{\centering}
    			%p{1in}
    			%p{3in}
    			%p{0.57\textwidth}
    		rlccc
    	}
    		\toprule
    		Parameter & Value & Estimator & CIL & CIU
    		\\
    		\midrule
    		
    		$\beta_s$ 
    		& \num{0.062552727}
    	    & \num{0.064248584} & \num{0.044765648}&
    	    \num{0.083731577}
    		\\
    		$\beta_a$ 
    		& \num{0.505426631}
    		& \num{0.504001573} & \num{0.481698536} &
    		\num{0.526304634}
    		\\
    	
    		$p$ 
    		& \num{0.305504967}
    		&\num{0.295644967} & \num{0.288265613}&
    		\num{0.303024437}
    		\\			
    		$\sigma$ & \num{1e-04}   & \num{9.940299E-005}
    		& \num{9.761672E-005}&
    		\num{1.011893E-005}\\
    		\bottomrule
    	\end{tabular}
    	\caption{Average and  95\% confidence interval of estimator of $\sigma$ and MLEs for $\beta_a,\beta_s$, and $p$.}
    	\label{table:MLE2}
    \end{table*}

    \section{Application to real data}\label{sec:application_real_data}
    \subsection{Data description and construction}\label{des_data}
    
    In this section, we use the dataset of new symptomatic and 
    confirmed COVID19 reported cases daily in Mexico City considering 
    that its population is $N=\num{26446435}$. The dataset contains 
    \num{47} records in the period March 10 to April 25 of 2020. We use 
    these records $I^o_s$ to construct the observations of the process 
    $I_s=I^{mx}_s:=I^o_s/N$ of the system \eqref{eqn:sde_model}.  
    Combining $I_s$ and Milstein scheme, we generate the
    others four processes of system \eqref{eqn:sde_model} using Algorithm \ref{Alg-ds}.  With the parameters of 
    \Cref{table:parametermodel}, $\sigma=1/100$ and $\Delta=1/1000$ Algorithm \ref{Alg-ds} works as follows.
    
    \begin{algorithm}[H]
  \begin{algorithmic}[1]
 \STATE Choose the 
    initial values of Table \ref{table:initial} for $E(0),I_a(0)$ and $R(0)$, $S(0)=1-E(0)-I_a(0) -R(0)-I^{mx}_s$, and make $n=0$.
    \STATE Generate $\Delta W\sim N(0,\Delta)$.
    \STATE $S(t_{n+1})=S(t_n)-(\mu+\beta_aI_a(t_{n})+\beta_s I_s^{mx}(t_n))\Delta S(t_{n})+(\mu+\gamma R(t_n))*\Delta-\sigma(1-S(t_n))\Delta W$
    \STATE $E(t_{n+1})=E(t_{n})-(\kappa+\mu)*\Delta E(t_{n})+\Delta*(\beta_sI_s^{mx}(t_n)+ \beta_aI_a(t_{n})S(t_n)-\sigma(E(t_n))\Delta W$
\STATE $I_a(t_{n+1})=I_a(t_{n})-p\kappa E(t_{n})\Delta I_a(t_{n})-(\alpha_a+\mu)I_a(t_{n})\Delta-\sigma I_a(t_{n})\Delta$
\STATE $R(t_{n+1})=1-S(t_{n+1}-E(t_{n+1})-I_a(t_{n+1})-I^{mx}_s(t_{n+1})$.
\STATE If $n<47$ make $n=n+1$ and go to 2. In other case stop.

\end{algorithmic}
  \caption{Construction of dataset.}\label{Alg-ds}
\end{algorithm}
    
    With this procedure, we have 47 records in the same observation 
    period for each of the five processes 
    that should fit to the stochastic model given by
    \eqref{eqn:sde_model}. In other words, we have observations of the
    processes of system \eqref{eqn:sde_model} at the times
    $t_0 = 0$,
    $t_1 = \num{0.001}$,
    $t_2 = \num{0.002}$,
    $\ldots$,
    $t_{46} = \num{0.046}$. 

\subsection{Estimation of parameters}\label{es_real}    
    
    We assume the following parameters as given:
$\gamma =1 /\num{365}$,
$\kappa = \num{0.196078}$,
$p = \num{0.5855049}$,
$\alpha_a = \num{0.167504}$ 
$\alpha_s = \num{0.0925069}$.
Thus, using the described methodology 
in \Cref{des_data} to generate data and applying 
\eqref{est_sigma},  we estimate $\sigma$ 
 and using the expressions  
\eqref{beta_hat_f}, \eqref{beta_hat_a_f}, \eqref{p_hat} we estimate  
 $\beta_s,\beta_a$, and $p$.

Next, we generate \num{1000} datasets using the data of Mexico City. 
\Cref{table:MLEREAL} gives the average of these estimators and the corresponding \num{95}\% confidence
interval. 
    \begin{table*}[tbh]
    	\centering
    	\begin{tabular}{%w
    			%>{\centering}
    			%p{1in}
    			%p{3in}
    			%p{0.57\textwidth}
    		rccc
    	}
    		\toprule
    		Parameter  & Estimator & CIL & CIU
    		\\
    		\midrule
    		
    		$\beta_s$ 
    		
    	    & \num{ 0.059159} & \num{0.002546}&
    	    \num{0.115772}
    		\\
    		$\beta_a$ 
    	
    		& \num{0.509925} & \num{0.378248} &
    		\num{0.641603}
    		\\
    	
    		$p$ 
    		
    		&\num{  0.582808} & \num{0.582326}&
    		\num{0.5832893}
    		\\			
    		$\sigma$ & \num{1.31701E-002}
    		& \num{8.32156E-003}&
    		\num{1.94434E-002}\\
    		\bottomrule
    	\end{tabular}
    	\caption{
    	    Average and  95\% confidence interval of estimator of 
    	    $\sigma$ and MLEs for $\beta_a,\beta_s$, and $p$.
    	}
    	\label{table:MLEREAL}
    \end{table*}
    
\begin{figure}[htb]
    \centering
    \includegraphics[width=0.8\textwidth, keepaspectratio]{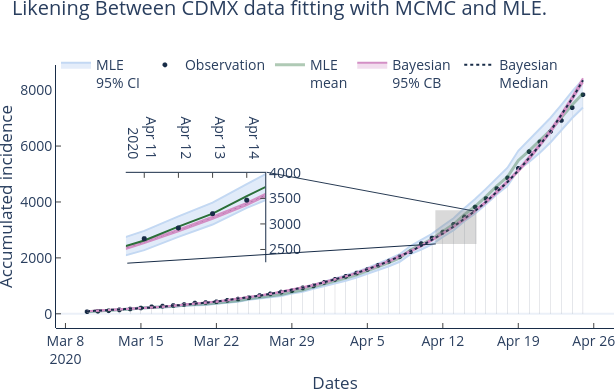}
    \caption{
        Comparison between the estimation of accumulated incidence of
        new symptomatic cases according to the Bayesian MCMC and our
        MLE approach. We refer the reader to 
        \url{https://chart-studio.plotly.com/~sauld/47} for a plot visualisation.%
    }
    \label{fig:real_data_fitting}
\end{figure}

\Cref{fig:real_data_fitting} shows the average and the \num{95} \%  
confidence interval of \num{1000} paths of the cumulative 
new symptomatic and confirmed COVID-19 reported cases from Mexico 
City.
%
% using the model \eqref{eqn:sde_model} generated using the described 
% methodology in Section \ref{des_data} and the estimators of 
% \Cref{table:MLEREAL} and the corresponding values for the Bayesian 
% estimate in the deterministic model. 
%
    Note that our estimation follows the given data profile 
in a closer way than the corresponding Bayesian calibration. 
Further, all observations drop in the corresponding confidence interval. 
So, this suggests that our proposed methodology and the stochastic 
extension improves the Bayesian fit of this dataset.
  
\subsection{Validation of the model}
 \begin{figure}[htb]
    \centering
    \includegraphics[scale=0.6]{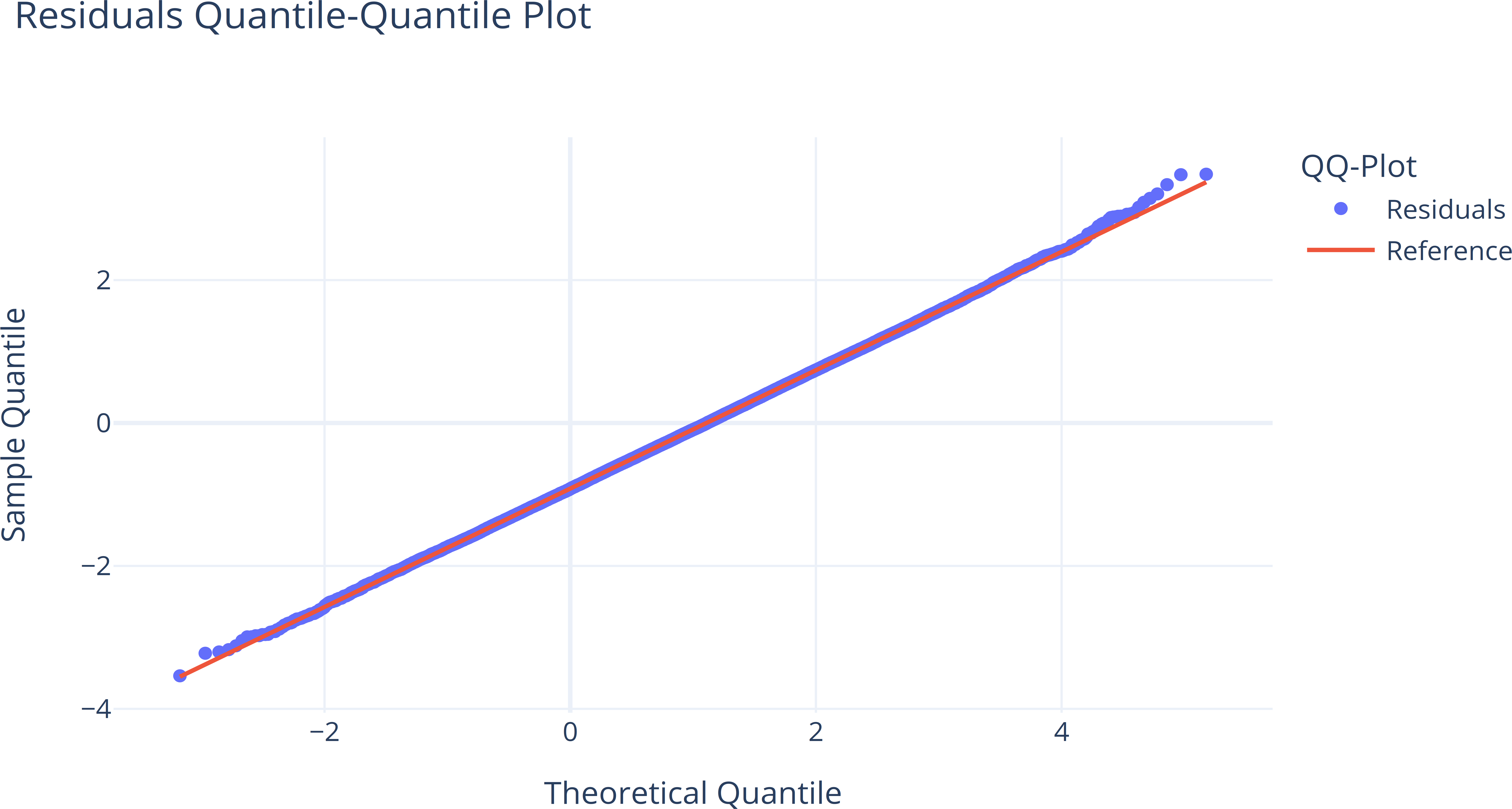}
    \caption{
        Quantile-Quantile plot of the numerical residuals.
        Here we plot the quantile for the error computed by Equation \eqref{Wt_increments}
        and a theoretical Gaussian distribution  distribution with zero mean and variance 1. 
    }
     \label{fig:fig_qq}
 \end{figure}
  
To validate our stochastic model, we compute the residuals corresponding 
to data of reported and confirmed COVID-19 cases of Mexico City.
Our target is to show that the residuals are Gaussian. If this is the case, 
then the noise $W$ in our stochastic model would be consistent with Brownian
motion.
 \\
Using equation \eqref{dlogS}, we define the increments 
\begin{equation}
    \begin{aligned}
        \Delta W(t_k)
        &:=
            \big[W(t_k)-W(t_{k-1})\big]
            \\ 
        &= -\frac{1}{\sigma}
            \Big[ 
                \log 
                \big(1-S(t_k) \big)
                    - \log\big(1-S(t_{k-1}) \big) 
            \Big]  - 
            \Big[ 
                \frac{\mu}{\sigma}+\frac{1}{2}\sigma 
            \Big]\Delta 
        \\
        &\quad +  
        \frac{\Delta }{\sigma}
        \Bigg(
            \Bigg[
                \frac{S(t_{k-1}) f_\beta(t_{k-1})}{1-S(t_{k-1})}
            \Bigg]
            - \gamma 
            \Bigg[
                \frac{R(t_{k-1})}{1-S(t_{k-1})} 
            \Bigg] 
        \Bigg),\label{Wt_increments}
    \end{aligned}
\end{equation}
where $\Delta=t_k-t_{k-1}$ and $f_\beta(t_{k})$ is the evaluation of the 
function $f_\beta$  in  time $t_{k}$; that is, 
\begin{align*}
    f_{\beta}(t_k) =& \beta_s I_s(t_k) + \beta_a I_a(t_k).
\end{align*}

    If the right-hand side of \eqref{Wt_increments} is Gaussian 
$\mathcal{N}(0,\Delta)$ for $k=1,\ldots, 46$ , then this will 
imply that the increments are Gaussian and also implies that the stochastic 
process $W$ is a Brownian motion, and therefore the stochastic model is 
well-fitted.

    We note that in expression \eqref{Wt_increments} there are 
four components of the stochastic processes, namely 
$ S, R, I_a, I_s$, but we only observed the component $I_s$ and the other
three processes had to be approximated using the methodology from the previous
sections.

    We use the parameter values of Section \ref{es_real}, estimators of 
\Cref{table:MLEREAL} and $\Delta=\num{0.001}$ 
into \eqref{Wt_increments}  and we get the residual 
$\Delta\widehat{W}(t_k)$. 
Then, we prove that  $\Delta\widehat{W}(t_k)$ are Gaussian with zero mean 
and finite variance, which leaves us to conclude that the model fits 
properly to the actual data. Indeed, \Cref{fig:fig_qq} shows a QQ-plot of 
standardised residual of \num{1000} paths of the real data versus a sample
of same size of standard normal random variable. 

Since an important part of the quantile frequency in 
\Cref{fig:fig_qq} follows the reference profile,
we can conclude that the residuals are Gaussian. 
This suggests that our stochastic model is well-fitted to the data.

    %\section{Numerical results}
    \section{Conclusions}
    
    We studied the MLE for some 
    key parameters of a stochastic SEIR model. 
    These parameters are important to study and quantify 
    uncertainty related to the stochastic nature of data.  
    By using synthetic data, we verified that our approximation 
    of the MLE is a good estimator for the infection transmission,
    symptomatic ratio and noise intensity parameters. 
    In the supplementary material, we provide a detailed
    proof for the Gaussian consistency of our
    estimators and other important properties.
    
        We
 then fitted  our stochastic model to a dataset of
    COVID-19 confirmed symptomatic cases from M\'exico City.
    Because this information is limited and incomplete in the 
    context of the regarding model, the database lacks
    of reliable information of the susceptible, exposed, 
    asymptomatic and recover classes, we implemented 
    the ideas as was explained in  Section \ref{des_data}.
    Thus, via simulation, we obtain confidence intervals 
    that capture more observations than the given MCMC outcome with its
    respective deterministic version, see \Cref{fig:real_data_fitting}. 
        
        One of our goals was to improve the uncertainty quantification 
    of COVID-19 data by a stochastic SEIR structure  based on It\^o SDEs. 
    Our simulations suggest that our MLE-estimators improve
    the variance description of a prescribed overfitting MCMC outcome. 
    Previous results confirmed this behavior in other applications, 
    see for example \cite{Chatzilena2019}. 
    Although our findings
     do not implicate 
    a general methodology, most of the presented techniques
    would be applicable to problems of the same sort.

        The estimators developed here capture the noisy 
    nature of the data with more precision. Furthermore, our simulations run in  much less 
    time with respect to the MCMC run with the STAN implementation. 
    However, our conclusions rely on a very particular input for the
    underlying outcome, and we depend on the algebraic form  of a 
    given model to construct an estimator. We have confirmed the 
    well-fitness of our stochastic model to real data, via error residual
    computing. However, it is necessary to explore our ideas in other 
    epidemic structures, and with other datasets and target parameters. 
    We should also develop more and stronger theories to construct 
    consistent and computable versions of our estimators.

% \paragraph{Strengths and weaknesses in relation to other studies, 
%    discussing important differences in results}
    
        As was discussed in the introduction, \citeauthor{PaGrGrMa} 
    reports the estimation of a  simple stochastic SIS structure that needs 
    only a scalar version of the model to approximate the likelihood. 
    There are more complex stochastic models similar to the one studied 
    here,  but they have not addressed estimation methods, see for instance \cite{faranda2020modeling} or  \cite{djordjevic2021two}. In contrast, 
    our results deal with an epidemic model of five non-linear coupled SDEs. Furthermore, we combine the stochastic model
    with real data for COVID-19 from Mexico City and, using a natural
    discretised version of the MLE, we estimate the parameters. 
    Therefore we fit the model to the actual data, and we obtain a 
    good behavior of the stochastic model.

        \citeauthor{Otunuga2021} also derives and calibrates a similar
    SEIR stochastic model for COVID-19 in \cite{Otunuga2021}. Their methodology 
    relies on in the Stratonovich stochastic calculus and new estimation scheme , which they 
    named 'lagged adapted generalized method of moments' (LLGMM). The LLGMM is a dynamic non-parametric method and seems to be a well-behaved method; however, the properties of the estimated parameters (e.g. consistency, bias, efficiency, etc.) are (up to our knowledge) still an open question.  Since we base our proposal in the It\^o framework,
    and the MLE method, the results of our
contribution
 are
 complementary.

One limitation of our work is that we have only one noise in all of the system. We then we pretend to extend the model in such manner that each process has an independent noise.  One
 open question is if such model could be susceptible to be studied with the methodology developed in this work, which means applying the MLE method. In addition, it will be necessary to study if the model is well-fitted to the actual data.

We also want to investigate if the  model studied in this manuscript (and a possible extension) could be applied to other type of diseases or phenomena. Another question is to determine, via simulation, the  minimum number of observations that ensure a nice approximation to the true parameters. Furthermore, this work missed a proof of the asymptotic normality of the estimators, which we guess that could
 be achieved by using the so-called Malliavin-Stein method. In addition, we want to study rates of convergence of the discretised version of the estimator to the continuous one. 
    \appendix
    \section{Proof of the consistency theorem}
        \label{two_theorems}

This section is devoted to present the proof of Theorem \ref{Consitency-th}.

    For the sake of completeness,  we first present a version of a 
classical Strong Law of Large Numbers that we use in getting our result.
 The proof of this particular version can be found, for instance, in Shiryaev \cite[Theorem IV.3.2]{ShiryaevBookProbability}). 

\begin{theorem}[Strong Law of Large Numbers]
    \label{SLLN}
Let $\xi_k, \ k\geq 1,$ be independent random variables with the following
properties:
\begin{itemize}
    \item $\E \xi_k=0$,\ $\E \xi_k^2>0$,
    \item There exist real numbers $c>0$ and $ \alpha\geq -1$ such that
    $$
    \lim_{k\to \infty} k^{-\alpha}\E \xi_k^2=c.
    $$
\end{itemize}
Then, with probability one,
$$
    \lim_{N\to \infty} \frac{\sum_{k=1}^N\xi_k}{\sum_{k=1}^N \E \xi_k^2}=0.
$$
If, in addition, $\E \xi_k^4 \leq c_1\Big(\E\xi_k^2\Big)^2$ for all $k\geq 1$,
with $c_1>0$ independent of $k$, then, also with probability one,
$$
    \lim_{N\to \infty} 
        \frac{\sum_{k=1}^N \xi_k^2}{\sum_{k= 1}^N \E \xi_k^2}=1.
$$
\end{theorem}

We now present the proof of the consistency of the MLE parameters.

\begin{proof}{\it (Of Theorem \ref{Consitency-th})} 
We first will get some estimates for $J_\star(T)$ and $J_{sa}(T)$. Observe that using Hypotheses \ref{set_hypotheses} we obtain
\begin{align*}
J_\star(T)&=\int_0^T \left(  \Big[ \frac{S(t) I_\star(t)}{ (1-S(t))} \Big]^2  +   \Big[ \frac{S(t) I_\star(t)}{ E(t)} \Big]^2 \right) dt\\
& < \int_0^T \left(  \Big[ \frac{S(t) I_\star(t)}{ (1-S(0))} \Big]^2  +   \Big[ \frac{S(t) I_\star(t)}{ E(0)} \Big]^2 \right) dt\\
&\le   \Bigg( \frac{1}{ (1-S(0))^2}   +   \frac{1}{ E^2(0)} \Bigg) 2T.
\end{align*}
In a very similar manner
\begin{align*}
J_{sa}(T)&= \int_0^T I_a(t) I_s(t) \left(  \Big[ \frac{S(t) }{ (1-S(t))} \Big]^2  +   \Big[ \frac{S(t) }{ E(t)} \Big]^2 \right) dt\\
& <  \Bigg( \frac{1}{ (1-S(0))^2}   +   \frac{1}{ E^2(0)} \Bigg) 2T.
\end{align*}
which implies that
\begin{align*}
-J_{sa}^2(T)&
> - \Bigg( \frac{1}{ (1-S(0))^2}   +   \frac{1}{ E^2(0)} \Bigg)^2  (2T)^2.
\end{align*}
At other hand, since $(1-S(t))^2<1$ and $E^2(t)<1$ then
$$
\frac{1}{(1-S(t))^2}>1,\quad \mbox{and } \quad \frac{1}{E^2(t)}>1   
 $$ 
then we have
\begin{align*}
J_\star(T)&=\int_0^T \left(  \Big[ \frac{S(t) I_\star(t)}{ (1-S(t))} \Big]^2  +   \Big[ \frac{S(t) I_\star(t)}{ E(t)} \Big]^2 \right) dt\\
& > \int_0^T \left(  \Big[ S(t) I_\star(t)\Big]^2  +   \Big[ S(t) I_\star(t)\Big]^2 \right) dt\\
&> 2  \Big[ S(T) I_\star(0)\Big]^2 T.
\end{align*}
For $J_{sa}$ we obtain a similar bound
\begin{align*}
J_{sa}(T)& =  \int_0^T I_a(t) I_s(t) \left(  \Big[ \frac{S(t) }{ (1-S(t))} \Big]^2  +   \Big[ \frac{S(t) }{ E(t)} \Big]^2 \right) dt\\
&>  \int_0^T I_a(t) I_s(t)  2S^2(t)  dt>   2T I_a(0) I_s(0)  S^2(T) 
\end{align*}
which implies
\begin{align*}
-J_{sa}(T)& <  -\Big( 2T I_a(0) I_s(0)  S^2(T) \Big)^2.
\end{align*}
From this estimations we get
\begin{align*}
4T^2\Bigg(  \Bigg( \frac{1}{ (1-S(0))^2}  & +   \frac{1}{ E^2(0)} \Bigg)^2 - S^4(T)I_s^2(0)I_a^2(0) \Bigg)> J_a(T)J_s(T)-J_{sa}^2(T)\\
& > 4T^2\Bigg( S^4(T)I_s^2(0)I_a^2(0)- \Bigg( \frac{1}{ (1-S(0))^2}   +   \frac{1}{ E^2(0)} \Bigg)^2 \Bigg)\\
& =: 4T^2 C_T
\end{align*}

and from this we obtain
\begin{align*}
\frac{1}{|J_a(T)J_s(T)-J_{sa}^2(T)|}&<\frac{1}{ 4T^2 C_T}
\end{align*}
where 
\begin{align*}
    C_T := 
        \Bigg|
            \Bigg( \frac{1}{ (1-S(0))^2} 
                + \frac{1}{ E^2(0)} 
            \Bigg ) ^ 2 - 
            S(T)^4  I_s(0)^2 I_a(0)^2 \Bigg|
\end{align*}

We now focus on the quantity $\hat{\beta_s}-\beta_{s,0}$, using \eqref{MLE-beta-beta0} we can write 
\begin{align*}
\hat{\beta_s}-\beta_{s,0} &= \frac{-\sigma}{J_s(T)J_a(T)- J_{sa}^2(T)} J_a(T) \mathcal{I}_1(T) + \frac{\sigma}{J_s(T)J_a(T)- J_{sa}^2(T)} J_{sa}(T) \mathcal{I}_2(T)\\
 &\le \frac{\sigma}{|J_s(T)J_a(T)- J_{sa}^2(T)|} J_a(T) \mathcal{I}_1(T) + \frac{\sigma}{|J_s(T)J_a(T)- J_{sa}^2(T)|} J_{sa}(T) \mathcal{I}_2(T)\\
&=: \Lambda_{1,1}+ \Lambda_{1,2},
\end{align*}
where 
\begin{align*}
 \mathcal{I}_1(T):= \int_0^T \Big[ \frac{S(t) I_s(t)}{ (1-S(t))} +  \frac{S(t) I_s(t)}{ E(t)} \Big]  dW(t), \\
  \mathcal{I}_2(T):= \int_0^T \Big[ \frac{S(t) I_a(t)}{ (1-S(t))} +  \frac{S(t) I_a(t)}{ E(t)} \Big]  dW(t).
\end{align*}

We have that 
\begin{align*}
\Lambda_{1,1}
 &\le \frac{\sigma 2T }{4T^2 C_T} \Bigg[\frac{1}{ (1-S(0))^2}  +   \frac{1}{ E^2(0)} \Bigg]\mathcal{I}_1(T) 
\end{align*}
 and 
 \begin{align*}
\Lambda_{1,2}
 &\le \frac{\sigma 2T }{4T^2 C_T}\Bigg[\frac{1}{ (1-S(0))^2}  +   \frac{1}{ E^2(0)} \Bigg] \mathcal{I}_2(T).
\end{align*}

We note that $\mathcal{I}_1(T)$ and $\mathcal{I}_2(T)$ are stochastic processes with zero mean and finite variance. Indeed, by the It\^o isometry we get, for instance for $\mathcal{I}_1(T)$ 
\begin{align*}
 \E\Big[\mathcal{I}_1(T)\Big]^2 = \int_0^T \E \Big[ \frac{S(t) I_s(t)}{ (1-S(t))} +  \frac{S(t) I_s(t)}{ E(t)} \Big]^2  dt
\end{align*}
which is finite. For $\mathcal{I}_2(T)$ is similar.

Define the sequence of real numbers $t_0=0$ and $t_k=T(1-\frac{1}{k})$ for $k\ge 1$. Thus, $t_k \uparrow T$ as $k \rightarrow \infty$. Moreover, 
$$
t_{k+1} - t_k = T (\frac{1}{k}-\frac{1}{k+1})= T \frac{1}{k(k+1)}< T \frac{1}{k}.
$$
Define the random variables
\begin{align*}
\xi_k := \int_{t_k}^{t_{k+1}} \Big[ \frac{S(t) I_\star(t)}{ (1-S(t))} +  \frac{S(t) I_\star(t)}{ E(t)} \Big]  dW(t).
\end{align*}
Thus, as before $\E\big[\xi_k  \big]=0$ and 
\begin{align*}
\E\big[\xi_k^2  \big] &= \E \int_{t_k}^{t_{k+1}} \Big[ \frac{S(t) I_\star(t)}{ (1-S(t))} +  \frac{S(t) I_\star(t)}{ E(t)} \Big]^2 dt \\
&=  \E \int_{t_k}^{t_{k+1}} \Big[ \frac{S^2(t) I_\star^2(t)}{ (1-S(t))^2} +  2 \frac{S^2(t) I_\star^2(t)}{ (1-S(t)) E(t)}  +\frac{S^2(t) I_\star^2(t)}{ E^2(t)} \Big] dt \\
&< \int_{t_k}^{t_{k+1}}  \Big[ \frac{1)}{ (1-S(0))^2} +  2 \frac{1}{ (1-S(0)) E(0)}  +\frac{1}{ E^2(0)} \Big] dt\\
&= \Big[ \frac{1)}{ (1-S(0))^2} +  2 \frac{1}{ (1-S(0)) E(0)}  +\frac{1}{ E^2(0)} \Big] (t_{k+1}-t_k)\\
&<  \Big[ \frac{1)}{ (1-S(0))^2} +  2 \frac{1}{ (1-S(0)) E(0)} + \frac{1}{ E^2(0)} \Big]\, \frac{T}{k}.
\end{align*}
Then, we have that the following limit is true
$$
\lim_{k\rightarrow \infty } k\, \E\big[\xi_k^2  \big]  < T \,  \Big[ \frac{1)}{ (1-S(0))^2} +  2 \frac{1}{ (1-S(0)) E(0)} + \frac{1}{ E^2(0)} \Big],
$$
This means that there exists $C>0$ such that 
$$
\lim_{k\rightarrow \infty } k^{-\alpha} \E\big[\xi_k^2  \big]  = C, \qquad \mbox{with } \alpha= -1.
$$
Then, by using the Theorem \ref{SLLN} we have that, with probability one,
\begin{equation}\label{convergence1}
\lim_{N \rightarrow \infty } \frac{\sum_{k=0}^N \xi_k }{\sum_{k=0}^N \E(\xi_k^2)  }= 0.
\end{equation}
We observe that 
$$
\sum_{k=0}^N \xi_k = \int_0^{t_{N+1}} \Big[ \frac{S(t) I_\star(t)}{ (1-S(t))} +  \frac{S(t) I_\star(t)}{ E(t)} \Big]  dW(t)
$$
and
\begin{align*}
\sum_{k=0}^N \E(\xi_k^2) & <  \Big[ \frac{1)}{ (1-S(0))^2} +  2 \frac{1}{ (1-S(0)) E(0)} + \frac{1}{ E^2(0)} \Big] \sum_{k=0}^N  (t_{k+1}-t_k)\\
& = \Big[ \frac{1)}{ (1-S(0))^2} +  2 \frac{1}{ (1-S(0)) E(0)} + \frac{1}{ E^2(0)} \Big]\, t_{N+1} =: A\, t_{N+1}.
\end{align*}
Then, from \eqref{convergence1} we deduce that 
$$
\frac{\int_0^{t_{N+1}} \Big[ \frac{S(t) I_\star(t)}{ (1-S(t))} +  \frac{S(t) I_\star(t)}{ E(t)} \Big]  dW(t)}{t_{N+1} A } \longrightarrow 0, \quad \mbox{when } N\rightarrow\infty.
$$
with probability one. This implies that $\Lambda_{1,1}+ \Lambda_{1,2} \rightarrow 0$ which consequently proves that $\hat{\beta_s}\rightarrow \beta_{s,0}$ with probability one. 

A similar procedure proves that $\hat{\beta_a}\rightarrow \beta_{a,0}$  with probability one.

We now focus on the proof of the consistency for $\hat p$. We rewrite \eqref{MLE_p} as
\begin{align}\label{MLE_p_rew}
\hat{p}_{ML}-p_0 &=  \Bigg( \frac{\sigma }{\E\big(J_3(T))} \int_0^T \Big[  -\frac{\kappa E(t)}{ I_a(t)} +
\frac{\kappa E(t)}{ I_s(t)} \Big] dW(t)\Bigg) \times \frac{\E\big(J_3(T))}{J_2(T) } ,
\end{align}
with $J_2$ as defined before and
$$
J_3(T):=  \int_0^T  \Big[ \frac{\kappa E(t)}{ I_s(t)} +  \frac{\kappa E(t)}{ I_a(t)}\Big]^2 dt.
$$
We will show that the first term in the right side of \eqref{MLE_p_rew} goes to zero with probability one and the second is bounded with probability one. As before, for the sequence $\{t_k\}_{k\ge 0}$ we define the random variables 
$$
\xi_k:=\int_{t_k}^{t_{k+1}} \Big[  -\frac{\kappa E(t)}{ I_a(t)} +
\frac{\kappa E(t)}{ I_s(t)} \Big] dW(t)
$$
and we have that $\E(\xi_k)=0$ and $\E(\xi_k^2)>0$. Moreover, we can have the following estimate for $\E(\xi_k^2)$:
$$
\E(\xi_k^2)< \Big[ \frac{\kappa }{ I_s(0)} +  \frac{\kappa }{ I_a(0)}\Big]^2 \frac{T}{k},
$$
and thus the following limit is satisfied
$$
\lim_{k\rightarrow \infty } k\, \E(\xi_k^2)< \Big[ \frac{\kappa }{ I_s(0)} +  \frac{\kappa }{ I_a(0)}\Big]^2 T,
$$
Therefore, there exists a constant $C>0$ such that $ \lim_{k\rightarrow \infty } k\, \E(\xi_k^2)=C$. Then, by the Theorem \ref{SLLN} we have that, with probability one,
\begin{equation}\label{convergence2}
\lim_{N \rightarrow \infty } \frac{\sum_{k=0}^N \xi_k }{\sum_{k=0}^N \E(\xi_k^2)  }= 0.
\end{equation}

This implies that
$$
\lim_{N \to \infty } 
    \dfrac{\int_0^{t_{N+1}} 
    \Big[
        -\dfrac{\kappa E(t)}{ I_a(t)} + 
        \dfrac{\kappa E(t)}{ I_s(t)} 
    \Big] dW(t) }{
        \int_0^{t_{N+1}}  
        \Big[
            \dfrac{\kappa E(t)}{ I_s(t)} + 
            \dfrac{\kappa E(t)}{ I_a(t)}
        \Big] ^ 2 dt } = 0
$$
meaning that the first term in \eqref{MLE_p_rew} goes to zero with probability one.

For the second term in \eqref{MLE_p_rew}, observe that
$$
\frac{\E\big(J_3(T))}{J_2(T) } \le  \frac{\E\big(J_2(T))}{J_2(T) }.
$$
Consider the random variable $\frac{J_2(T)}{\E\big(J_2(T))}$. We have that 
$$
\E\Bigg(\frac{J_2(T)}{\E\big(J_2(T))} \Bigg)=1.
$$
Furthermore, the random variable $J_2(T)$ is positive with probability one and then has a moment $\E\big(J_2(T))\ge 0$, thus for $n\in\N$ 
\begin{align*}
\P\Bigg(\frac{\E\big(J_2(T))}{J_2(T)}\ge n   \Bigg) =  \P\Bigg(  \frac{J_2(T)}{\E\big(J_2(T))}\le \frac{1}{n} \Bigg) = 1 - \P\Bigg(  \frac{J_2(T)}{\E\big(J_2(T))}\ge \frac{1}{n} \Bigg).
\end{align*}
We note that the quotient 
$
\frac{J_2(T)}{\E\big(J_2(T))} $
is non negative with probability one, then we deduce that, when $n \rightarrow \infty $,
\begin{align*}
 \P\Bigg(  \frac{J_2(T)}{\E\big(J_2(T))}\ge \frac{1}{n} \Bigg)  \rightarrow  1,
\end{align*}
therefore,
\begin{align*}
    \P\Bigg(
        \frac{\E\big(J_2(T))}{J_2(T)}\ge n   \Bigg) \rightarrow  0 \quad \mbox{ when } 
        \to \infty.
\end{align*}
This implies that the random variable $\dfrac{\E\big(J_2(T)\big)}{J_2(T)}$ is bounded  with 
probability one. That is $\hat{p}_{ML}\rightarrow p_0$  with probability one, which 
completes the proof of the theorem.
 \end{proof}

    \section{Some calculations}
        \label{Ap-1}
            It is possible to prove (see supplementary material) that
\begin{equation}
    \begin{aligned}
        \label{beta_hat_f}
        \hat{\beta}_{s,ML}  
            = &   
            \dfrac{1 }{J_s J_a - J_{sa}^2} 
            \Bigg[ 
                J_a(T) \int_0^T  \dfrac{S(t) I_s(t)}{ (1-S(t))} 
                \Big[
                    d
                    \big(
                        \log (1-S_t) 
                    \big)
                \Big] 
            \\
            & 
            + J_a(T) 
        \int_0^T  
            \dfrac{S(t) I_s(t)}{ (1-S(t))} 
            \Big[
                \Big(  
                    \mu + 
                    \frac{\gamma  R}{ (1-S)} + 
                    \tfrac{1}{2} 
                    \sigma^2 
                \Big) dt 
            \Big] 
        \\
        &
            + 
            \textbf{} J_a(T) 
            \int_0^T
                \frac{S(t) _s(t)}{ E(t)} 
                d(\log (E_t)) + J_a(T) 
                \int_0^T 
                \frac{S(t) I_s(t)}{ E(t)} 
                \Big(
                    \mu  
                    +
                    \frac{1}{2}
                    \sigma^2 
                \Big) dt 
                +\
                \kappa
                J_a(T) 
                \int_0^T
                    \frac{S(t) I_s(t)}{ E(t)} 
                    dt
        \\
        &
            - J_{sa}(T) 
                \int_0^T
                    \frac{S(t) I_a(t)}{ (1-S(t))}  
                    d\big(\log (1-S_t) \big) -  J_{sa}(T) 
                \int_0^T
                    \frac{S(t) I_a(t)}{ (1-S(t))} 
                    \Big(
                        \mu + \frac{\gamma  R}{ (1-S)} + \tfrac{1}{2} \sigma^2
                    \Big) dt
        \\
        & \ 
            -   J_{sa}(T) \int_0^T   
                \frac{S(t) I_a(t)}{ E(t)}  d(\log (E_t)) -
                J_{sa}(T) 
                    \int_0^T
                        \frac{S(t) I_a(t)}{ E(t)}
                        \Big(
                            \mu + 
                            \frac{1}{2} 
                            \sigma^2 
                        \Big) dt 
        \\
        & - \kappa J_{sa}(T)
            \int_0^T
                \frac{S(t) I_a(t)}{ E(t)} dt
            \Bigg] 
    \end{aligned} 
\end{equation}
and

\begin{equation}
\label{beta_hat_a_f}
    \begin{aligned}
        \hat{\beta}_{a,ML} &=
            \frac{1 }{J_s J_a - J_{sa}^2}
            \Bigg[
                - J_{sa}(T) 
                \int_0^T 
                \frac{S(t) I_s(t)}{ (1-S(t))}
                \Big[ d\big(\log (1-S_t) \big)
            \Big] 
            \\
        &\, 
        -  J_{sa}(T) 
        \int_0^T  
            \frac{S(t) I_s(t)}{ (1-S(t))} 
            \Big[ 
                \Big(
                    \mu 
                    + \frac{\gamma  R}{ (1-S)} 
                    + \tfrac{1}{2} \sigma^2 
                \Big) dt 
            \Big]
        \\
        & \,  
            -\textbf{}
                 J_{sa}(T) 
                 \int_0^T
                    \frac{S(t) I_s(t)}{ E(t)}
                    d(\log (E_t)) 
            -\textbf{} 
                J_{sa}(T) 
                \int_0^T
                    \frac{S(t) I_s(t)}{ E(t)}
                    \Big(
                        \mu 
                        + \frac{1}{2}
                        \sigma^2 
                    \Big) dt  
            -  \, 
                \kappa 
                    J_{sa}(T) 
                \int_0^T
                    \frac{S(t) I_s(t)}{ E(t)} dt 
        \\
        &\, 
            + J_{s}(T) 
            \int_0^T
                \frac{S(t) I_a(t)}{ (1-S(t))}
                \Big[
                    d\big(\log (1-S_t) \big)
                \Big]
        \\
        &\, 
            + J_{s}(T) 
            \int_0^T
            \frac{S(t) I_a(t)}{ (1-S(t))}
            \Big[
                \Big(
                    \mu 
                    + \frac{\gamma  R}{ (1-S)} 
                    + \tfrac{1}{2} \sigma^2
                \Big) dt 
            \Big]
        \\
            & \,   
            +\textbf{} J_{s}(T) 
            \int_0^T
                \frac{S(t) I_a(t)}{ E(t)}
                d(\log (E_t))
                +\textbf{} 
                J_{s}(T) 
                \int_0^T
                       \frac{S(t) I_a(t)}{ E(t)} 
                       \Big(
                            \mu +
                            \frac{1}{2} \sigma^2
                       \Big) dt   \,    
                       + \kappa  J_{s}(T)   
                       \int_0^T
                        \frac{S(t) I_a(t)}{ E(t)} dt
                    .
    \end{aligned}
\end{equation}

Moreover, 
\begin{equation}
    \begin{aligned}
        \label{p_hat}
        \hat{p}_{ML}  
            = &  
            \frac{1}{J_2(T)} 
            \Bigg[  
                \kappa^2 
                \int_0^T 
                    \frac{E^2(t) }{I_S^2(t)} 
                dt 
                - 
                \kappa 
                \int_0^T 
                    \frac{E(t) }{
                        I_S(t)
                    } d \big(\ln (I_S(t)) \big)
                    + 
                    \kappa 
                    \int_0^T 
                    \dfrac{E(t)}{I_a(t)} d\big(\ln (I_a(t)) \big)   
            \\
            & -  
                \kappa (\alpha_S+\mu) 
                \int_0^T 
                    \dfrac{E(t) }{I_S(t)} 
                dt 
                - 
                \dfrac{1}{2}\kappa \sigma^2 
                \int_0^T  
                \dfrac{E(t) }{I_S(t)} dt 
            \\
                &
                + \kappa (\alpha_a+\mu) 
                \int_0^T \frac{E(t) }{I_a(t)} dt 
                + 
                \dfrac{1}{2}
                \kappa \sigma^2 
                \int_0 ^ T 
                \dfrac{E(t)}{I_a(t)} dt
                \Bigg].
    \end{aligned}
\end{equation}

        \bibliographystyle{rss}
%        \bibstyle{rss}
        \bibliography{References.bib}
     %   \bibliographystyle{rss}
         %\bibstyle{rss}

\end{document}